\documentclass[a4paper,12pt]{article}

\usepackage{epstopdf}
\usepackage{caption}
\usepackage{graphicx,color,amsmath}

\begin{document}

\title{Oblique wrinkles}

\author{
Melania Carfagna$^{1}$, Michel Destrade$^{2,3}$, \\ Artur L. Gower$^{2,4}$, Alfio Grillo$^1$\\[12pt]
$^{1}$Dipartimento di Scienze Matematiche  G.L. Lagrange, \\
Politecnico di Torino, \\
Corso Duca degli Abruzzi, 24 Torino, Italy;\\[6pt]
$^{2}$School of Mathematics, Statistics and Applied Mathematics,\\
 NUI Galway, University Road, Galway, Ireland;\\[6pt]
$^{3}$School of Mechanical and Materials Engineering, \\
University College Dublin, Belfield, Dublin 4, Ireland;\\[6pt]
$^{4}$School of Mathematics, University of Manchester, \\
Oxford Road, Manchester, M13 9PL, United Kingdom.}

\date{}

\maketitle


\bigskip\bigskip\bigskip


\begin{abstract}

We prove theoretically that when a soft solid is subjected to an extreme deformation, wrinkles can form on its surface at an angle that is oblique to a principal direction of stretch. 
These oblique wrinkles occur for a strain that is smaller than the one required to obtain wrinkles normal to the direction of greatest compression. 
We go on to explain why they \color{black} will probably \color{black} never be observed in real-world experiments.

\end{abstract}


\newpage


\section{Introduction}


\begin{figure}[t!]
\centering
\includegraphics[width=\textwidth]{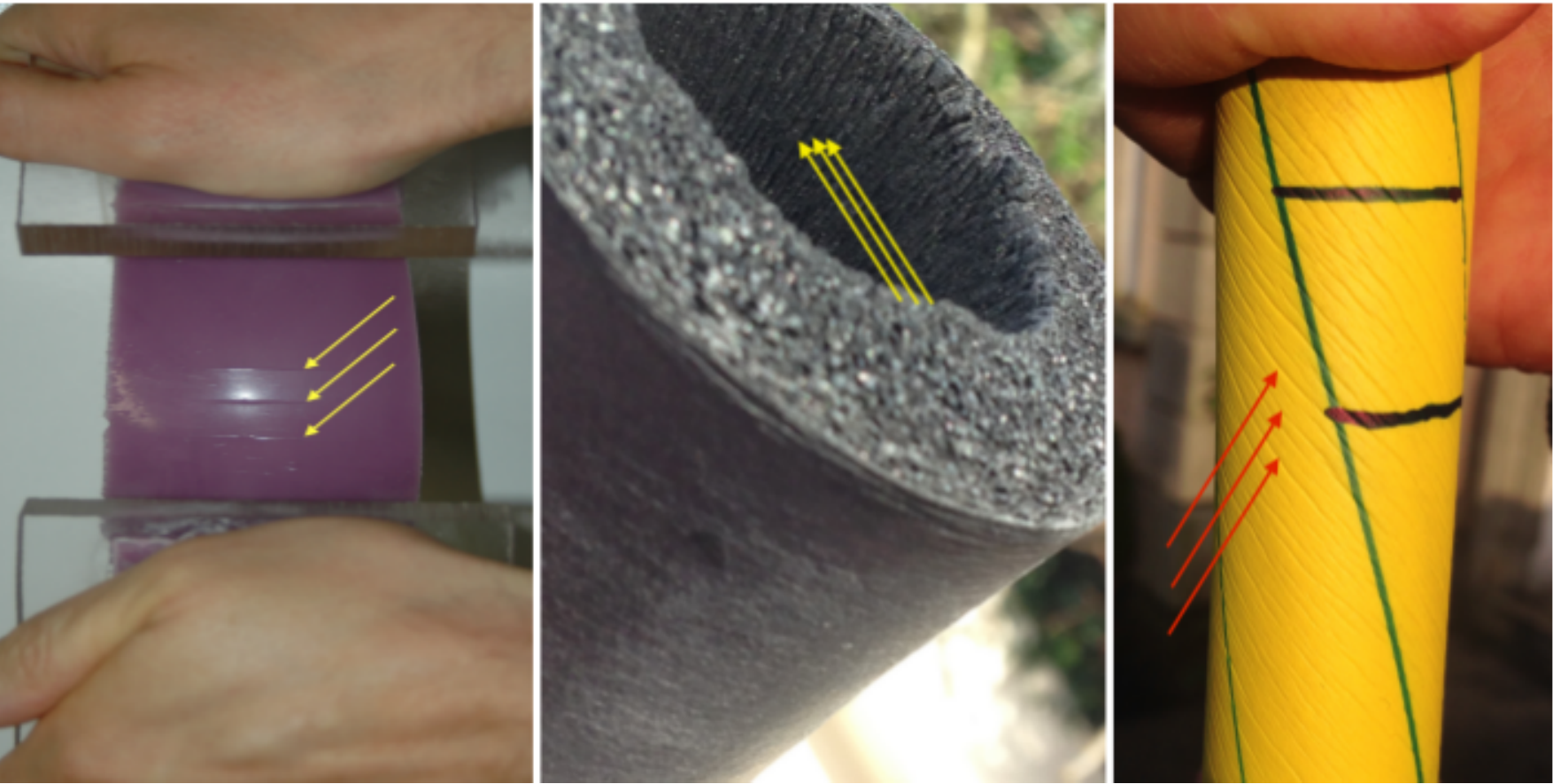}
\caption{Creases appearing in the large deformation of soft matter, as indicated by arrows. Left: Large bending of a silicone block with two ends glued to plastic plates, on which bending moments are applied; Middle: Everted tube of pipe insulation material; Right: Twisted cylinder, on which axial and circumferential lines were drawn prior to the twist. Notice how all creases are aligned with a principal direction of stretch.}
\label{fig:Figure_1}
\end{figure}

\begin{figure}[t!]
\centering
\includegraphics[width=\textwidth]{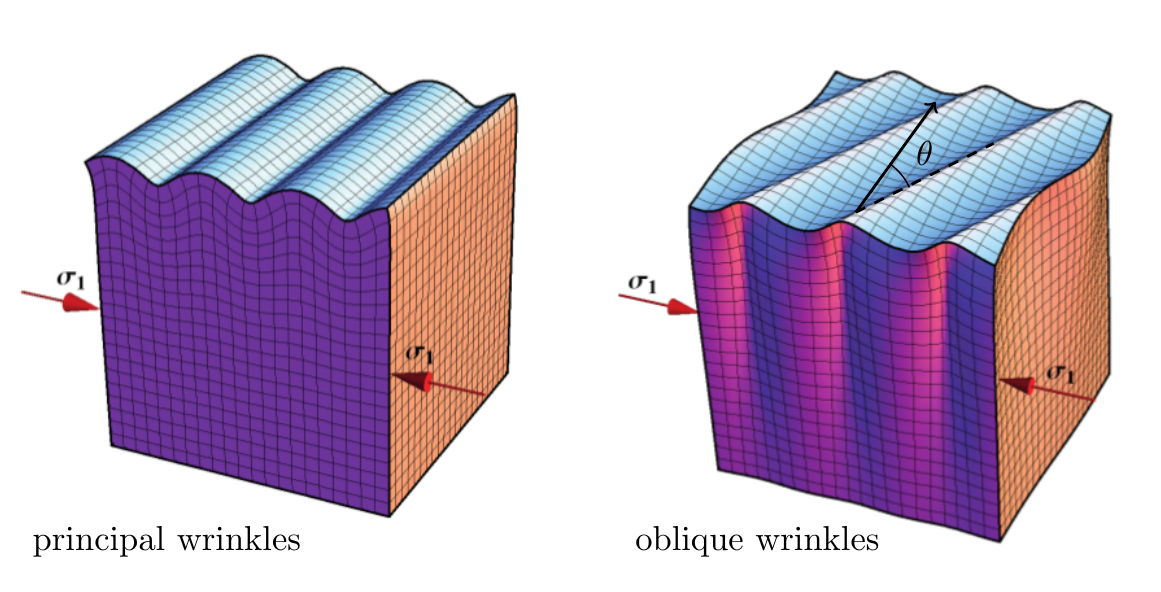}
\caption{Graphical representation of  \textit{principal} (left) and \textit{oblique} (right) wrinkles on the free surface of a compressed semi-infinite soft solid subject to a uni-axial compression with unique Cauchy stress component $\sigma_1$.}
\label{fig:Figure_2}
\end{figure}

When a soft solid is subjected to an extreme deformation, its surface (or a part of it) eventually buckles, provided it has not ruptured before. Guided by physical intuition and by experimental evidence (see for example Figure \ref{fig:Figure_1}), we expect  the formation of wrinkles or creases to be arranged orthogonally to the direction of greatest contraction, and thus aligned with a principal direction of deformation.
However, as we show in this paper, the mathematical equations modelling the development of such surface instabilities do not necessarily predict that they should be such \textit{principal} wrinkles (see Figure \ref{fig:Figure_2} on the left).
In fact, we find that for the simplest boundary value problem there is, i.e. that of a deformed semi-infinite solid, the theory predicts that \textit{oblique} wrinkles  (see Figure \ref{fig:Figure_2} on the right) should appear on the free surface prior to the principal wrinkles. We present these results in Section \ref{sec:Sec2}, where we study the theoretical predictions of wrinkle orientation for the most general model of isotropic, incompressible third-order elasticity soft solid or equivalently, the Mooney-Rivlin model. 
These two equivalent models provide a good mathematical description of isotropic homogeneous soft solids such as rubber, silicone, or gel, subjected to finite deformations.
We establish that, for a certain range of material parameters, the preferred direction of wrinkle formation is not a principal direction of deformation, and that the \textit{obliquity angle} can be as much as $33^\circ$, depending on the mode of pre-deformation.

In Section \ref{sec:Sec3} we report experimental attempts to uncover oblique wrinkles on the surface of a homogeneous block of gelatine subjected to a shear-box deformation \cite{AGower2014}. 
The search of such occurrence was completely unsuccessful. 
First of all we did not manage to generate actual sinusoidal wrinkles on the block, but we always obtained \emph{creases}, which are indeed expected to occur prior to the wrinkles \cite{Huth_and_Cao_2012,Huth_and_Cao_2016}.
Moreover, the creases were not oblique, but aligned with the long diagonal of the shear-box, itself a direction of principal stretch.

We moved on to tests with a slightly different set-up, allowing us to finally observe the formation of sinusoidal wrinkles instead of creases. 
For that set-up, we let the free surface of the gelatine dry overnight to a certain extent and form a stiff film on top of a softer substrate.
It is then well known \cite{Huth_and_Cao_2012} that for such a layered structure, sinusoidal wrinkles occur first, and not creases (provided the stiffness contrast is large enough); however, we observed that the wrinkles always appeared in a principal direction of deformation. 

In Section \ref{sec:Sec4}  we elucidate why oblique wrinkles do not appear experimentally, even in the case of a coated half-space.
Using the Stroh formalism \cite{Stroh1962} to predict their onset, we find that oblique wrinkles should appear, indeed, but only as long as the overlaying film is at most two times stiffer than the substrate. 
If it is stiffer than that, then the first wrinkles to appear are the principal ones according to our calculations. 
However, Hutchinson and Cao  \cite{Huth_and_Cao_2012} used post-buckling Finite Element simulations \color{black} of neo-Hookean solids \color{black} to show that creases dominate wrinkles as long as the film is less than about ten times stiffer than the substrate.
Hence, oblique wrinkles, although predicted by the theory, will \color{black} probably \color{black} never be captured in an experiment.


\section{Oblique wrinkles on a half-space}
\label{sec:Sec2}


To model isotropic, homogeneous, incompressible \textit{soft solids}, we use a strain energy density $W$ of the Mooney-Rivlin type. It is linear in $I_1=\mathrm{tr}(\mathbf{C})$ and $I_2=\mathrm{tr}(\mathbf{C}^{-1})$, where $\mathbf{C}$ is the right Cauchy-Green deformation tensor, 
\begin{align}\label{MR_Energy}
W=\tfrac{1}{4}\mu[(1-\beta)(I_1-3)+(1+\beta)(I_2-3)],
\end{align}
where $\mu>0$ is the shear modulus, and $\beta$ is a material parameter such that $-1\le\beta\le 1$. Note that these inequalities ensure the strong ellipticity of the incremental equations of equilibrium \cite{DestradeScott}. At $\beta=-1$, $W$ recovers the so-called neo-Hookean material (linear in $I_1$) and at $\beta=1$, the so-called Extreme-Mooney \cite{Shield} material (linear in $I_2$).

\color{black}
As shown by Rivlin and Saunders \cite{RiSa51}, when we expand the Mooney-Rivlin strain energy density in powers of the Green-Lagrange strain $\mathbf{E}=(\mathbf{C}-\mathbf{I})/2$, and neglect terms of order higher than cubic, \color{black} we find that (\ref{MR_Energy}) is equivalent to the most general model of isotropic, incompressible, third-order elasticity
\begin{align}\label{TO_Energy}
W=\mu_0\mathrm{tr}(\mathbf{E}^2)+\tfrac{1}{3}A\,\mathrm{tr}(\mathbf{E}^3),
\end{align}
where $\mu_0>0$ is the second-order Lam\'e coefficient and $A$ is the third-order Landau constant. 
The connection between those constants is  \cite{Destrade2010b} $\mu_0=\mu$, $A=-\mu(\beta+3)$, which puts bounds on $A$ in order to ensure strong ellipticity of the incremental equations: $-4\mu_0\le A\le -2\mu_0$. 

\begin{figure}[t!]
\centering
\includegraphics[width=2.6in]{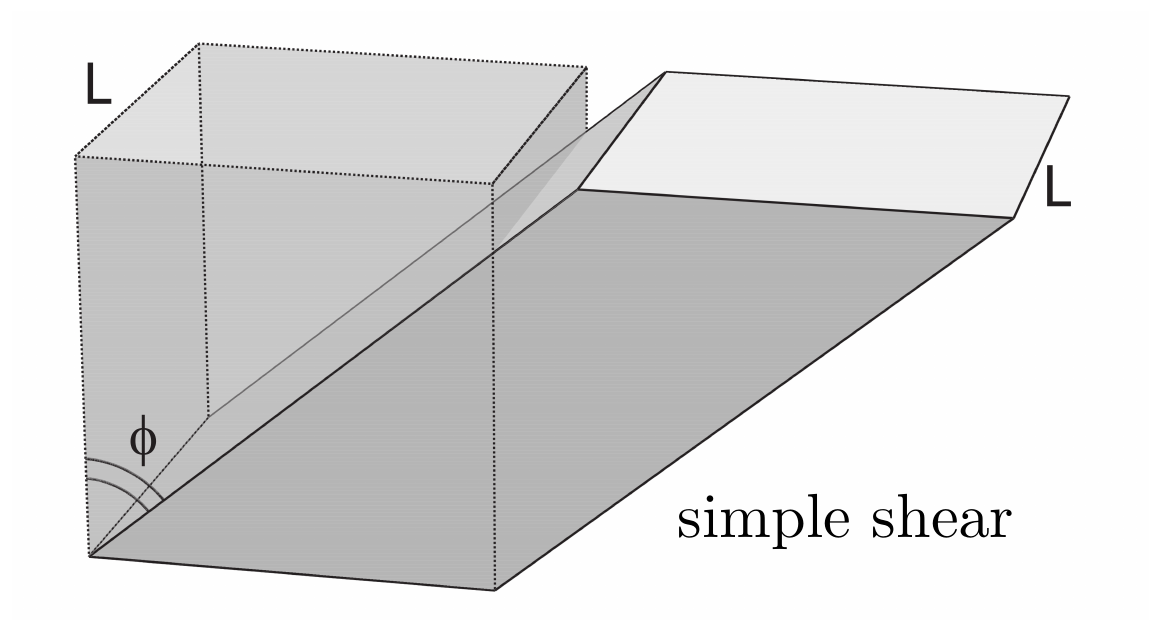}
\includegraphics[width=2.6in]{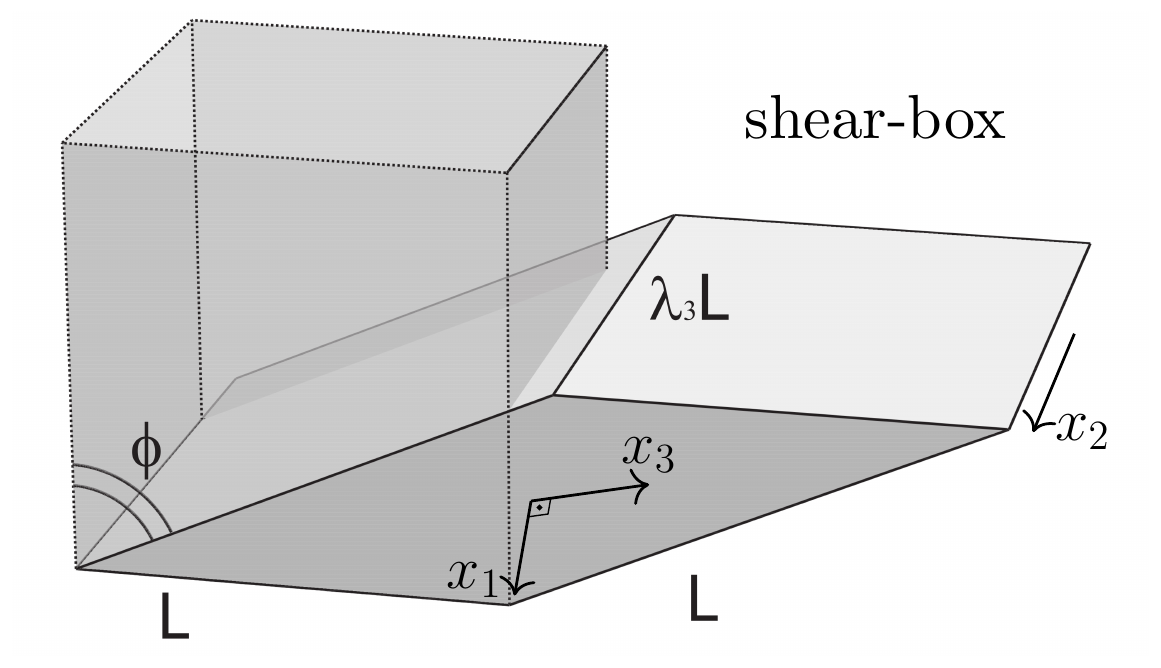}
\caption{Two shearing deformations and their tilting angle $\phi$ of a soft solid initially occupying a cube of side $L$. Left: \emph{Simple shear}, where the top and the bottom faces move on parallel planes. Right: \emph{Shear-box}, where the initial front and back square faces become rhombic. In the latter case, two of the Eulerian principal directions are aligned  with the diagonals of the rhombi.}
\label{fig:Figure_3}
\end{figure}

To model \textit{finite strains}, we consider homogeneous deformations of a semi-infinite solid. 
We place ourselves in the Cartesian coordinate system $(x_1,x_2,x_3)$ of axes aligned with the Eulerian principal directions (along the eigenvectors of $\mathbf{b}$, the left Cauchy-Green deformation tensor). 
We denote the eigenvalues of $\mathbf b$ by $\lambda_1^2$, $\lambda_2^2$ and $\lambda_3^2$, and call $\lambda_1=\lambda<1$ the principal stretch of contraction, $\lambda_2$ the principal stretch along the normal to the free surface, and $\lambda_3$ the third principal stretch. Because of the incompressibility constraint $\det \mathbf{b} = 1$, we know that $\lambda_3=(\lambda_1\lambda_2)^{-1}$. 
The homogeneous deformation is maintained by the application of a constant Cauchy stress $\boldsymbol \sigma$, and we take the stress component along the direction $x_2$ to be zero, i.e., $\sigma_{22}=0$, as the boundary of the half-space is assumed to be free of traction.
We will consider the following archetypes of deformation,
\begin{itemize}
\item[(i)] Uniaxial compression: $\lambda_1=\lambda$, $\lambda_2  = \lambda^{-1/2}$, $\lambda_3 = \lambda^{-1/2}$;
\item[(ii)] Plane strain: $\lambda_1 = \lambda$, $\lambda_2 = \lambda^{-1}$, $\lambda_3 = 1$;
\item[(ii)] Simple shear: $\lambda_1 = \lambda$, $\lambda_2 = 1$, $\lambda_3 = \lambda^{-1}$;
\item[(iv)] Shear-box: $\lambda_1 = \lambda$, $\lambda_2 = 1/(\lambda\sqrt{2-\lambda^2})$, $\lambda_3= \sqrt{2-\lambda^2}$.
\end{itemize}
and we will use $\lambda$ as our \textit{parameter of bifurcation}, calling it $\lambda_\mathrm{cr}$ when the threshold for wrinkles is reached. 
Note that other quantities can be used, such as the material compressive strain \cite{Huth_and_Cao_2012} $\varepsilon_\mathrm{cr}=1-\lambda_\mathrm{cr}$, or the angle of tilting $\phi$ for shearing deformations, see Figure \ref{fig:Figure_3}. 
For simple shear (iv), $\phi$ is given by $\tan\phi=\lambda^{-1}-\lambda$ \cite{Chad}, and for the shear-box deformation, by $\cos\phi=\lambda\sqrt{2-\lambda^2}$ \cite{Stolz}.

We denote by $\theta$ the \textit{obliquity angle}, i.e, the angle between the wavefront of the wrinkles and the direction of the largest stretch. 
At $\theta=0^\circ$, the wrinkles are normal to the direction of greatest compression, while at $\theta\ne 0^\circ$, they are \textit{oblique}, see Figure \ref{fig:Figure_2}.

For \textit{principal wrinkles}, the wavefront is aligned with $x_3$ and varies sinusoidally in the $x_1$-direction, while the amplitude decays exponentially with $x_2$. 
For the Mooney-Rivlin model, the bifurcation criterion for principal wrinkles is independent of the material constants $\mu$ and $\beta$ (or $\mu_0$ and $A$).
It reads as
\begin{equation}\label{eq:Flavin1}
\left(\frac{\lambda_1}{\lambda_2}\right)_\mathrm{cr}=\sigma_0 \simeq 0.2956,
\end{equation} 
where $\sigma_0$ is the real root of the cubic 
\begin{equation}\label{eq:cubic}
\sigma^3+\sigma^2+3\sigma-1=0.
\end{equation}
This result can be traced back  to Biot \cite{Biot}, Flavin \cite{Flavin}, and Green and Zerna \cite{Green}. 
In Table \ref{tab:Table_1}, we list the critical stretches and strains corresponding to the modes of deformation (i)-(iv) for principal wrinkles and for oblique wrinkles on an Extreme-Mooney material. \begin{table}[!h]
\centering
\caption{Critical stretches and strains for principal wrinkles in archetype modes of deformation, together with the associated angle of tilting for shear deformations, and the corresponding original reference (first five columns).
For oblique wrinkles, the critical stretches, strains, tilting angles and obliquity angles are reported for the Extreme-Mooney solids (last four columns).}
\label{tab:Table_1}
\begin{tabular}{lllllllll}
\hline
 &$\lambda_\mathrm{cr}$&$\varepsilon_\mathrm{cr}$ &$\phi_\mathrm{cr}$& Ref. & $\lambda_\mathrm{cr}^\mathrm{oblique}$ & $\varepsilon_\mathrm{cr}^\mathrm{oblique}$  & $\phi_\mathrm{cr}^\mathrm{oblique}$ &$\theta^\mathrm{oblique}$  \\
\hline
Uni-axial & 0.444 & 56\% &  & \cite{Biot} &0.537 &46 \%  & & $34^\circ$\\ 
Plane strain & 0.544 & 46\% &  & \cite{Biot} & 0.562 & 44 \%& & $29^\circ$\\
Simple shear & 0.296 & 70\% &$72^\circ$& \cite{Destrade2008} & 0.522 & 48 \%& $54^\circ$& $33^\circ$ \\
shear-box&0.471& 53\% &$51^\circ$&\cite{Stolz} & 0.539 & 46 \% &  45$^\circ$ & $33^\circ$
\\\hline
\end{tabular}
\vspace*{-4pt}
\end{table}

Destrade et al. \cite{Destrade2005} found an explicit secular equation for surface waves in deformed Mooney-Rivlin materials, from which the bifurcation criterion here can be found by taking the wave speed to be zero. We used this explicit secular equation to predict the formation of wrinkles (it is too long to reproduce here). 
In parallel, we also used the bifurcation criterion based on $\mathbf{z}$, the surface impedance matrix \cite{biryukov, Shuv00, Fu05, DeCN09}, which is a simple version of a method we will use for layered media in Section~\ref{sec:Sec4}. It reads
\begin{equation}
\mathrm{det} \, \mathbf{z}=0,
\end{equation}
where $\mathbf{z}$ is the Hermitian, positive semi-definite solution of the algebraic Riccati equation
\begin{equation}\label{eq:Riccati1}
\mathbf{z}\mathbf{N}_2\mathbf{z}- \text i\mathbf{z}\mathbf{N}_1+ \text i\mathbf{N}^\dag_1\mathbf{z}+\mathbf{N}_3=\mathbf{0}.
\end{equation}
Here $\mathbf{N}_1$, $\mathbf{N}_2$, $\mathbf{N}_3$ are the $3\times 3$ blocks of the $6\times6$ Stroh matrix (see the appendix), and the superscript $^\dag$ denotes the Hermitian conjugate.
To solve this equation numerically we rely on Riccati solvers found in Mathematica and Matlab, which give the positive definite solution. The results from this method were the same (within machine precision) as those given by the explicit secular equation found in \cite{Destrade2005}.
There are three short analytical expressions of the bifurcation criterion in some special cases. 
\begin{figure}[ht!]
\centering
\includegraphics[width=0.5\textwidth]{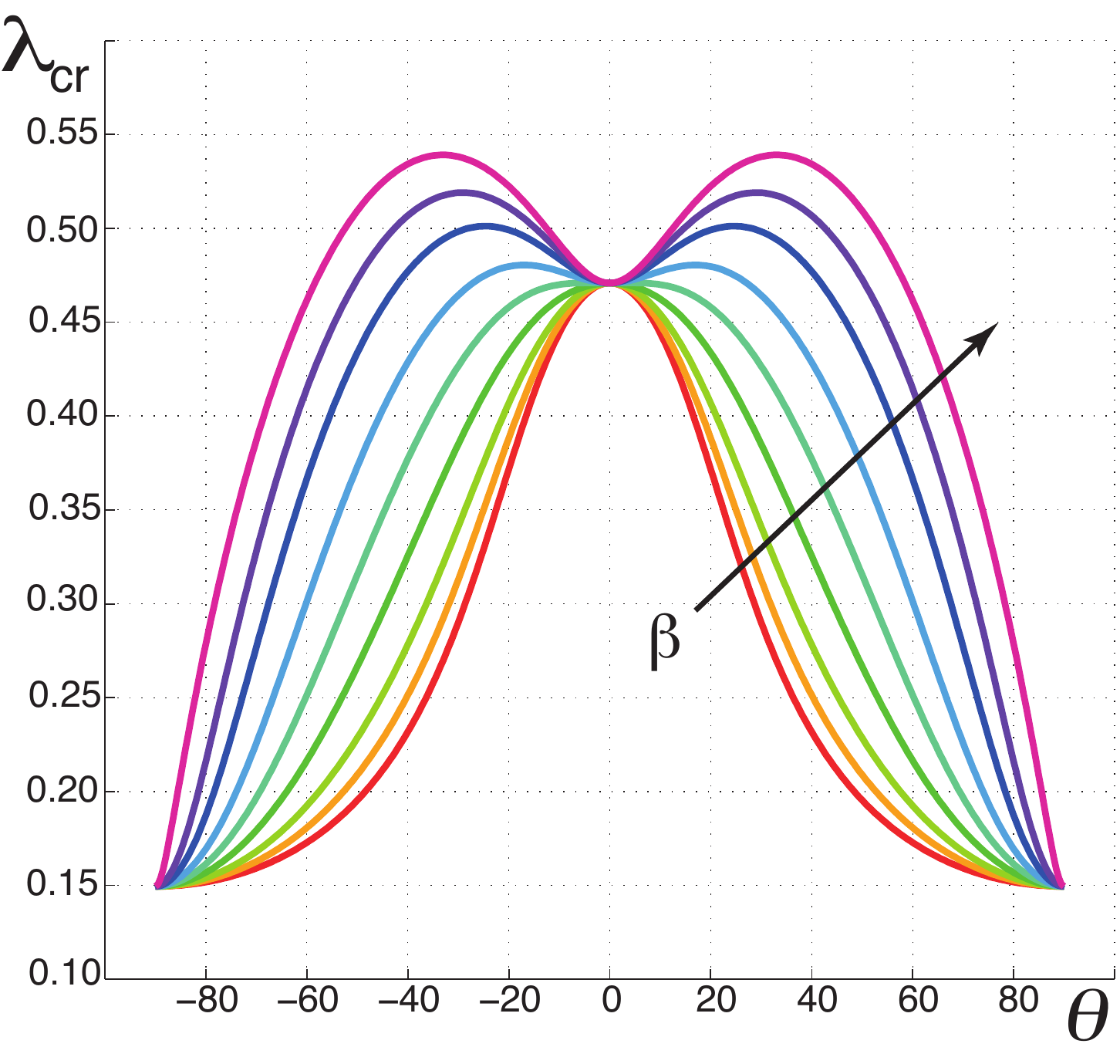}\quad
\raisebox{0.06\textwidth}{\includegraphics[width=0.47\textwidth]{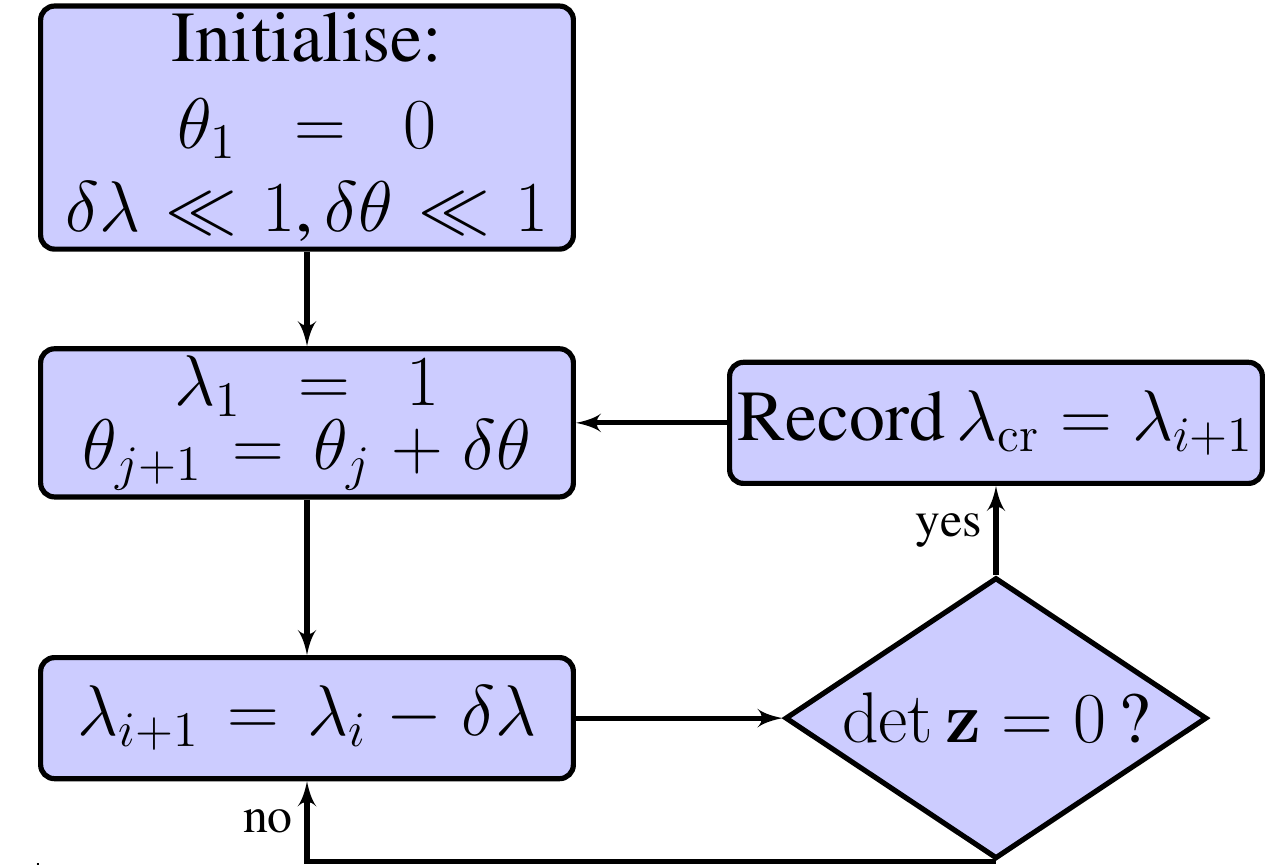}}
\caption{Left: Dependence of the critical compressive stretch $\lambda_\mathrm{cr}$ on the obliquity angle $\theta$, for different values of $\beta$ in the shear-box compression test. As $\beta$ varies from $\beta=-1$ (neo-Hookean material) to $\beta=1$ (Extreme-Mooney material) [$\beta = -1.0, -0.6, -0.3, 0, 0.3, 0.6, 0.8, 1.0$, see arrow], the maximum of the plot shifts from $\theta=0^\circ$ (principal wrinkles appearing at  53\% strain) to $\theta \simeq 33^\circ$ (oblique wrinkles, appearing at an earlier strain of 46\%).
Right: Flow-chart for the corresponding numerical procedure.}
\label{half-space-shear-box}
\end{figure}

For the neo-Hookean material (i.e., $\beta=-1$ in (\ref{MR_Energy}), or $A=-2\mu_0$ in (\ref{TO_Energy})), Flavin \cite{Flavin} found the following bifurcation criterion
\begin{align}\label{eq:Flavin2}
\lambda_1^2\lambda_3^2(\lambda_1^2\cos^2\theta+\lambda_3^2\sin^2\theta)=\sigma_0^2.
\end{align}

For the Extreme-Mooney material (i.e., $\beta=1$ in (\ref{MR_Energy}), or equivalently $A=-4\mu_0$ in (\ref{TO_Energy})), we find the following explicit bifurcation criterion 
\begin{align}\label{eq:Flavin3}
\sigma_0^4+\sigma_0^3+\lambda_1^2\lambda_3^2(\lambda_1^4\lambda_3^4-\lambda_3^2-\lambda_1^2)\sigma_0(\sigma_0+1)+4\lambda_1^6\lambda_3^6=0,
\end{align}
with $\sigma_0$ now given by Equation (\ref{eq:Flavin2}). 
Using Mathematica, the reals roots of (\ref{eq:Flavin3}) can be obtained, and then substituted into Equation (\ref{eq:Flavin2}) to get an explicit buckling criterion.

Finally, for a wrinkle-front aligned with the long diagonal of the shear-box ($\theta = 0^\circ$), we have again \textit{principal} wrinkles, and Equation (\ref{eq:Flavin1}) applies.

\color{black}
For Cases (i)-(iv) \color{black} we find that there is a range of values for $\beta$ starting at $\beta=-1$ (neo-Hookean solid), where principal wrinkles appear first, and another range up to $\beta=1$ (Extreme-Mooney solid), where we expect oblique wrinkles to appear. Figure \ref{half-space-shear-box} illustrates this occurrence for the shear-box deformation (Case (iv)), in which we see that principal wrinkles are expected when $-1\le \beta\le -0.3$, and oblique wrinkles when $-0.3\le \beta\le 1$. 
For the Extreme-Mooney material, the obliquity angle is $\theta \simeq 33^\circ$, and the critical stretch is $\lambda_\mathrm{cr}=0.539$,  compared to $\lambda_\mathrm{cr}=0.471$ for principal wrinkles.
We focused on this deformation because of its ease of experimental implementation (see next section), but the other deformations yielded a similar behaviour, with more or less marked differences between the values of $\lambda_\mathrm{cr}$ for principal  wrinkles and for oblique wrinkles, as summarised in Table \ref{tab:Table_1}.

In general, the numerical procedure to find the critical values of the compressive stretch and of the corresponding obliquity angle is straightforward and robust. 
After non-dimensionalisation, the critical values turn out to be independent of $\mu$ (or $\mu_0$), and so we can plot $\lambda_\mathrm{cr}$ against $\theta$ for a given solid characterised by the single material parameter $\beta$ (or the ratio $A/\mu_0$).


\section{Experimental buckling with the shear-box deformation}
\label{sec:Sec3}


To build a shear-box we assembled four acrylic plates $(10\;\mathrm{cm}\times10\;\mathrm{cm}\times1.0\;\mathrm{cm})$ into a cube with four hinged edges.
We then prepared 12 different gels, using commercial gelatine in several forms: solid leafs, powders, or concentrated cubes to be dissolved in water. We varied the concentrations, from that prescribed by the manufacturers to three times the normal concentration.
We poured about $300\mathrm{ml}$ of gelatine into the shear-box and let it set into a homogeneous solid phase, either at room temperature or at refrigerated temperature, from 4 to 12 hours depending on the specimen. 

\begin{figure}[ht!]
\centering
\includegraphics[width=0.49\textwidth]{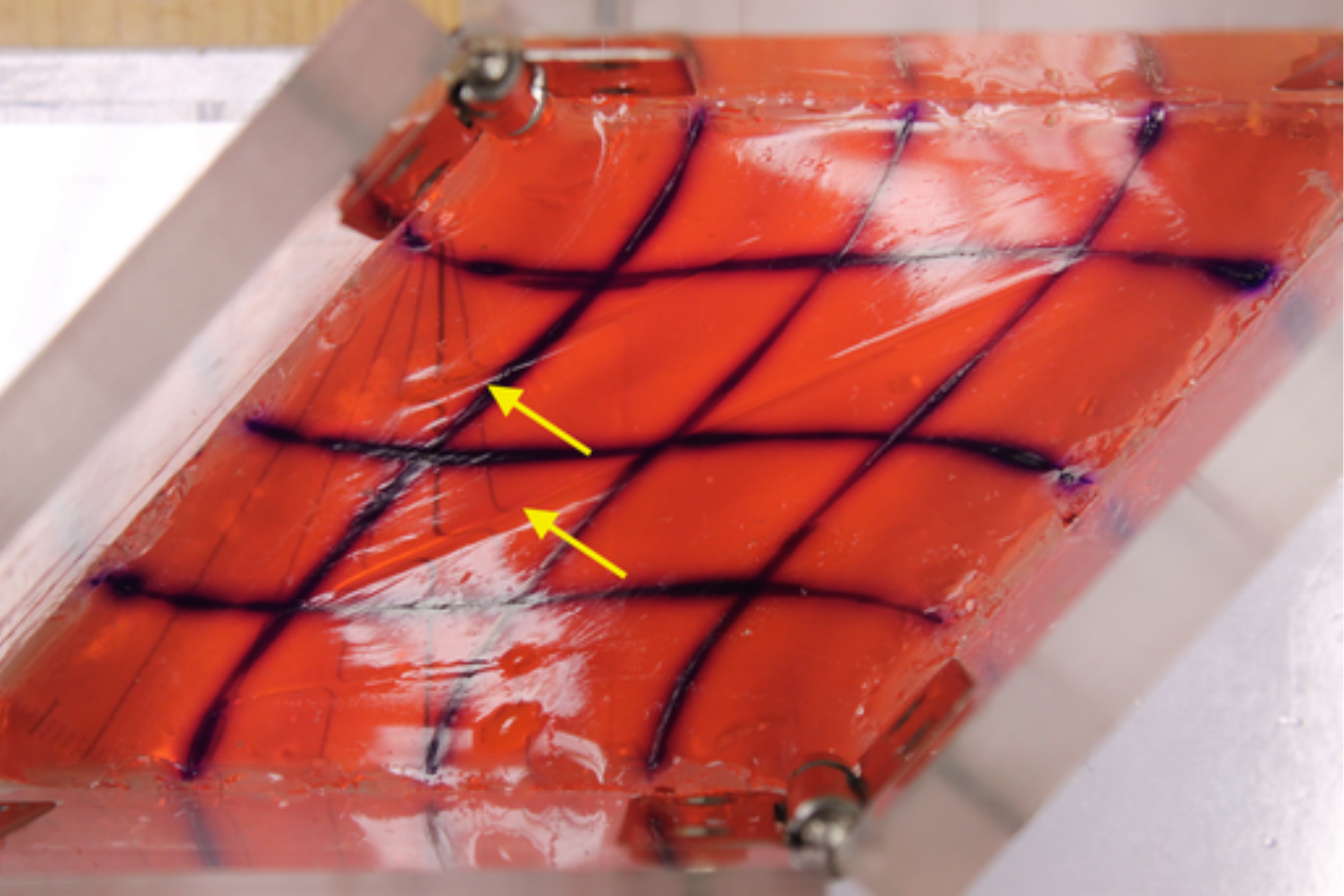}
\includegraphics[width=0.49\textwidth]{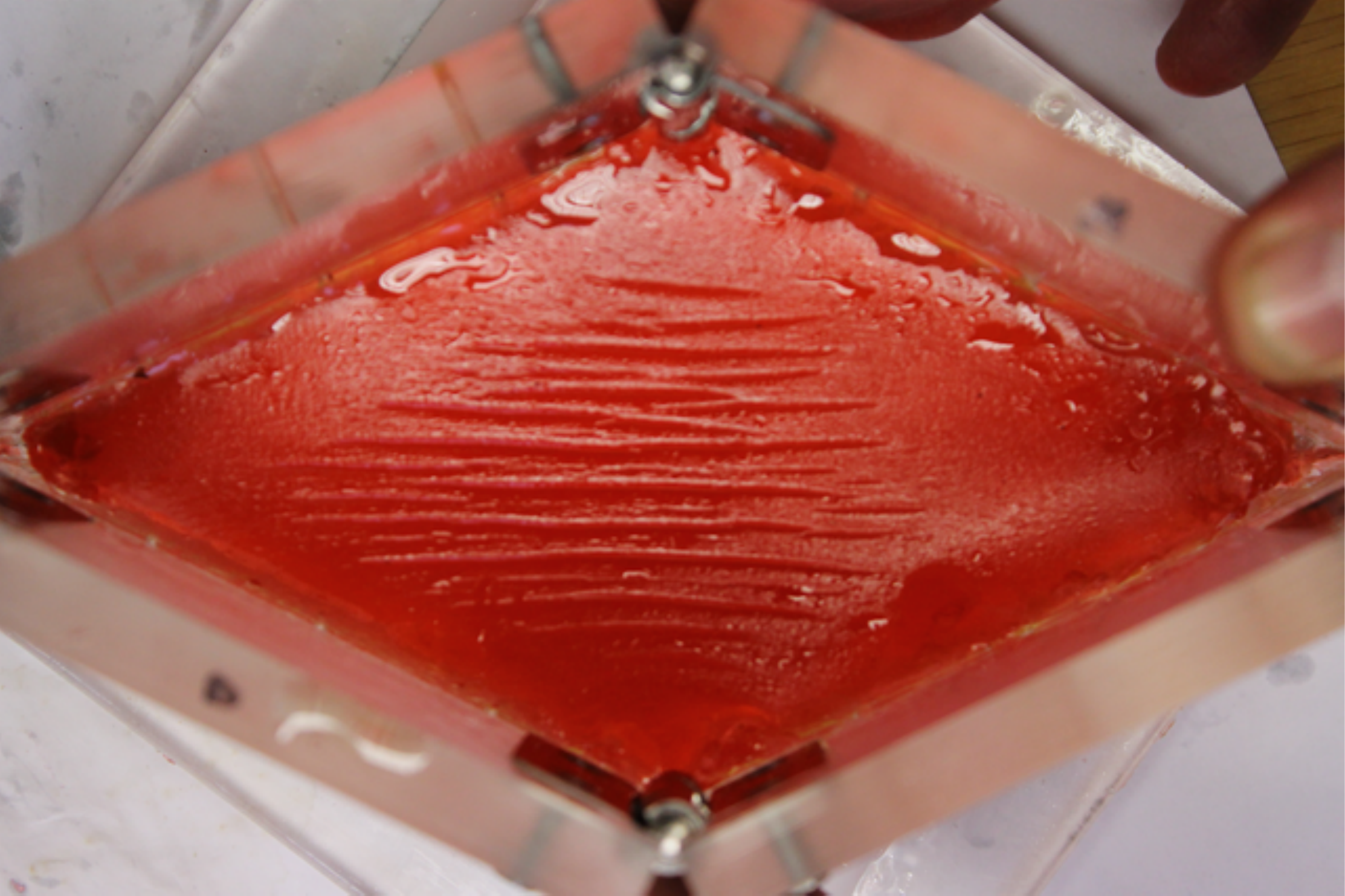}
\caption{Left:  Freshly solidified gelatine tends to exhibit surface creases instead of sinusoidal wrinkles when subject to an extreme deformation, see arrows. Right: Here the gelatine was left overnight to solidify and a stiffer layer/skin formed on the surface on top of the soft substrate; when deformed the coated block produced sinusoidal wrinkles.}
\label{creases-wrinkles}
\end{figure}

The gelatine block was then subjected to a shear-box deformation (Case (iv)), simply by applying manual pressure on two opposite hinged edges of the box (see Figure \ref{fig:Figure_3} on the right for a schematic representation of the deformed state).
\color{black} To ensure that the solid deformed homogeneously, we injected olive oil between the sides of the block and the walls to enhance the gliding.
This precaution lead to a central area where it was reasonable to consider that the deformation  was homogeneous, i.e. where initially parallel lines remained parallel in the current configuration, see Figure \ref{creases-wrinkles} on the left.
We carried on the deformation manually \color{black}
 until the free surface buckled as noticed by the naked eye, see examples on Figure \ref{creases-wrinkles}. 
The corresponding critical tilting angle $\phi$ was recorded.

Experimentally, we noted that we did not observe the formation of small-amplitude sinusoidal wrinkles but of creases instead, see Figure \ref{creases-wrinkles} on the left. 
Moreover the creases were always aligned with the long diagonal of the shear-box, and thus not oblique. 
We measured that the creases appeared at $\theta=38^\circ\pm2^\circ$, irrespective of the concentration and type of gelatine used for the gel sample. 

Hence we conclude that the theoretical predictions of Section \ref{sec:Sec2} were not verified experimentally. 
The main problem is that of crease formation trumping wrinkles; this was to be expected as creases have been predicted to appear much earlier than wrinkles on homogeneous blocks. 
For instance, Hong et al.\cite{Hong09} showed through Finite Elements simulations that creases appear on a  neo-Hookean half-space deformed in plane strain at a 35\% compressive strain, so that the 46\% required for sinusoidal wrinkles is never attained.

\begin{figure}[ht!]
\centering
\includegraphics[width=0.49\textwidth,  height=0.25\textwidth]{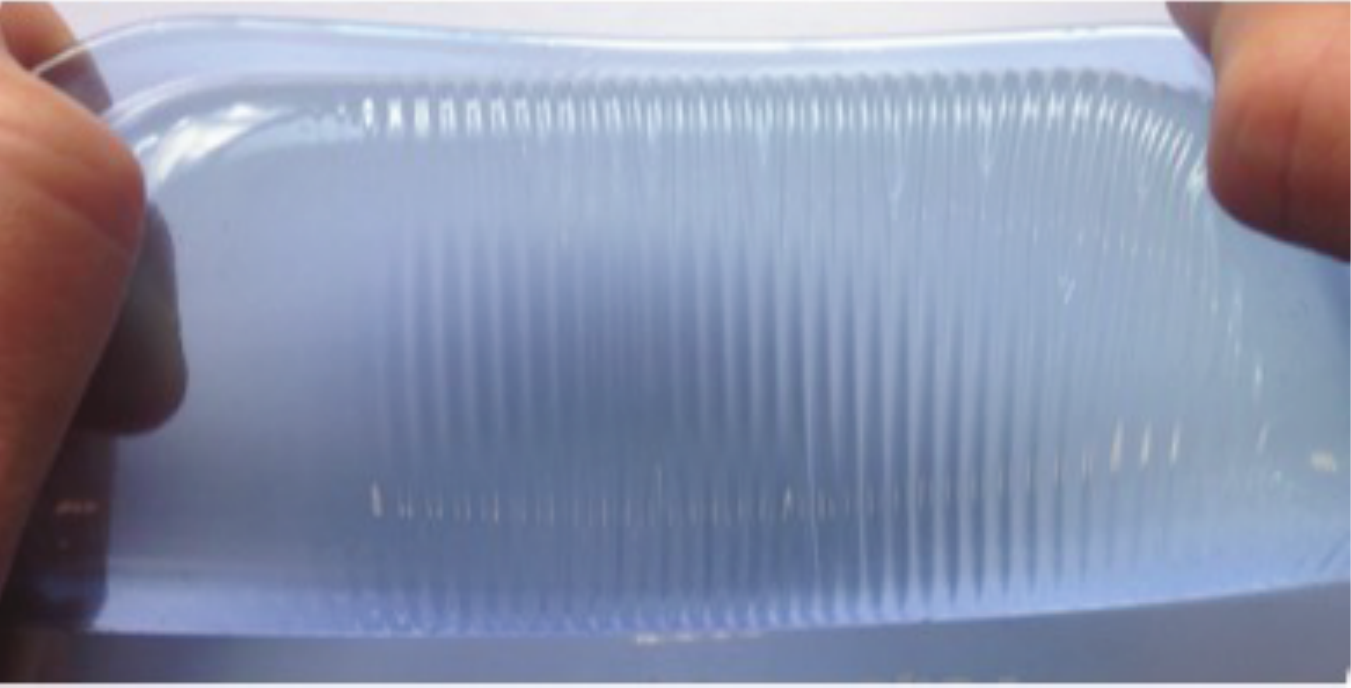}
\includegraphics[width=0.49\textwidth, height=0.25\textwidth]{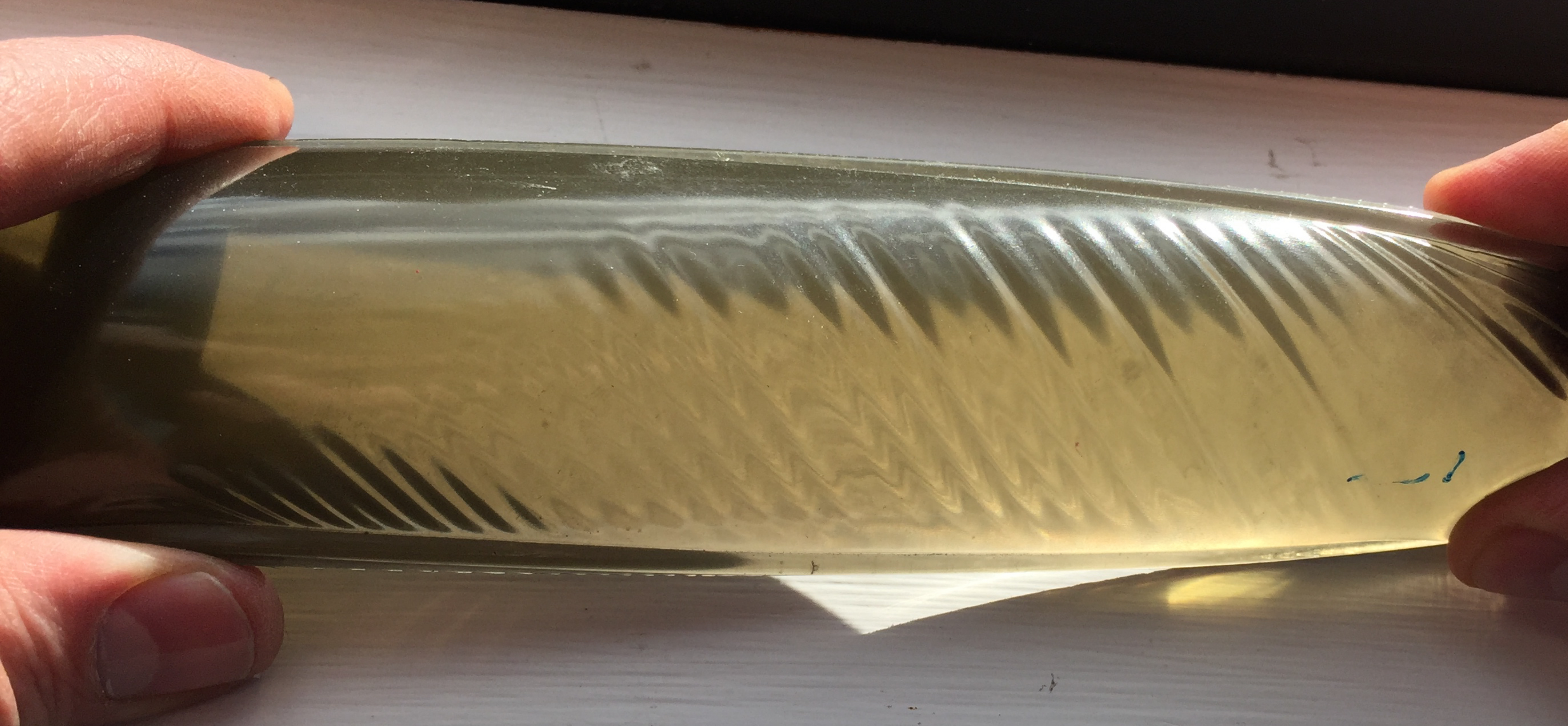}
\caption{Formation of regular sinusoidal wrinkles on wrist supports made of gels enclosed by a stiff film casing, by bending (left) or torsion (right).}
\label{wrinkles}
\end{figure}

To bypass this problem we focused instead on wrinkles appearing when a soft semi-infinite solid is coated with a thin stiff layer.
Those are then remarkably stable and their onset is well predicted by incremental analysis; they develop into nonlinear patterns only with further compression, see Cao and Hutchinson \cite{Huth_and_Cao_2012}  or \color{black}Jin et al. \cite{Jin15} \color{black} for details. 
Figure \ref{wrinkles} displays experimental evidence of the formation of regular sinusoidal wrinkles for bending and torsion deformations.

Experimentally we let the gelatine blocks dry out overnight at refrigerated temperatures, so that a skin formed on their surface due to surface dehydration during the longer cooling process.
When deformed in the shear-box, the buckling surface now exhibited regular, sinusoidal wrinkles, at least in the center of the block, where the homogeneous deformation was taking place, see Figure \ref{creases-wrinkles} on the right.
\color{black}
We noted that the wavelength of the wrinkles was much smaller than the dimensions of block, and thus that the half-space idealisation was justified for the modelling
\color{black}(for an analysis of the influence of the finite depth of the block, see the recent work by Jin et al. \cite{Jin15}).
\color{black}
However, oblique wrinkles were not observed, but principal wrinkles instead, again at a titling angle of $38^\circ \pm 2 ^\circ$ (corresponding to a strain of $\varepsilon_\text{cr} = 38\%$).
In the next section we explain why this absence of experimental oblique wrinkles is indeed in agreement with theoretical predictions.

\section{Wrinkles on a coated half-space}
\label{sec:Sec4}



\subsection{Layered structure}



We consider that the top surface of the gelatine block (in contact with air) has dried out and stiffened faster than the rest of the enclosed block during the solidification process.
We model the resulting heterogeneous structure as a composite material: a thin film of thickness $2h$, say, in the region $-h \le x_2 \le h$, in bonded contact with a substrate in the region $h \le x_2 \le \infty$.
The strain energy of each solid is expanded up to third order of Green strain as
\begin{equation}
 W_\mathrm{f} = \mu_\mathrm{0f} \, \text{tr}\left(\mathbf E^2\right) + \tfrac{1}{3}A_\mathrm{f} \, \text{tr}\left(\mathbf E^3\right), \qquad 
W_\mathrm{s} = \mu_\mathrm{0s} \, \text{tr}\left(\mathbf E^2\right) + \tfrac{1}{3}A_\mathrm{s}  \, \text{tr}\left(\mathbf E^3\right),
\end{equation}
respectively. 
Now consider the mechanical response of these two solids to common modes of deformation such as uni-axial tension under Cauchy tensile stress $T$ or simple shear under Cauchy shear stress $S$.
Then, up to second order in the strain, we have for the film and for the substrate 
\begin{align}
& T_\mathrm{f} = 3 \mu_\mathrm{0f} \, \varepsilon + 3(\mu_\mathrm{0f} + A_{\mathrm{f}}/4)\varepsilon^2, &&
S_\mathrm{f}= \mu_\mathrm{0f} \, \varepsilon, \notag \\
& T_\mathrm{s} = 3 \mu_\mathrm{0s} \, \varepsilon + 3(\mu_{0\mathrm{s}} + A_{\mathrm{s}}/4)\varepsilon^2, &&
S_\mathrm{s} = \mu_\mathrm{0s} \, \varepsilon, 
\end{align}
respectively, and similarly for other modes of deformation.
(Here, the measure of strain $\varepsilon$ is in turn the elongation $1 - \lambda$ (from Case (i)) and the amount of shear $\lambda^{-1} - \lambda$ (from Case (iii)).) 
A consistent way to ensure that the response of the film is always `stiffer' than that of the substrate is to take
\begin{equation} \label{stiffness-law}
\mu_\mathrm{0f} = \Gamma \mu_\mathrm{0s}, \qquad 
A_\mathrm{f} = \Gamma A_\mathrm{s}, 
\end{equation}
where $\Gamma>1$ is defined as the stiffness ratio. 
For the constants of the Mooney-Rivlin material (\ref{MR_Energy}), this translates as
\begin{equation}
\mu_\mathrm{f} = \Gamma  \mu_\mathrm{s}, \qquad \beta_\mathrm{f} = \beta_\mathrm{s}.
\end{equation}
Then the film is always stiffer than the substrate, in the sense that the magnitude of its response to any mechanical stress is $\Gamma$ times that of the substrate.


\subsection{Principal wrinkles}


For our layered structure, the formulas we use below can readily be deduced by specialising the results of \cite{ShEv02, Dest07} to the present context.
We write the bifurcation criterion for principal wrinkles as
\begin{equation} \label{crit}
\text{det}\left(\Gamma \, \mathbf z_{\mathrm{f}} - \mathbf z_{\mathrm{s}}\right) = 0,
\end{equation}
where  $ \mathbf z_{\mathrm{f}}$, $\mathbf z_{\mathrm{s}}$ are non-dimensionalised surface impedance matrices for the film and the substrate, respectively.
The substrate impedance is given explicitly by  $\mathbf z_{\mathrm{s}} = -\text i \mathbf{BA}^{-1}$, where
\begin{equation}\label{eq:AB}
 \mathbf A = \begin{bmatrix}
1 & \lambda_1^2\lambda_2^{-2}\\
\text i & \text i  \lambda_1\lambda_2^{-1}
\end{bmatrix},
\qquad
 \mathbf B = 
\begin{bmatrix}
\text 2\text i & \text i \lambda_1\lambda_2^{-1}(1+\lambda_1^2\lambda_2^{-2}) \\
-1 - \lambda_1^2\lambda_2^{-2} & -2 \lambda_1^2\lambda_2^{-2}
\end{bmatrix},
\end{equation}
and as before, $\lambda_1$ is the principal strech of contraction and $\lambda_2$ is the principal stretch along the normal to the surface.
The film impedance is given by $\mathbf z_\mathrm{f} =- \text i \mathbf M_3 \mathbf M_1^{-1}$, where $\mathbf M_1$, $\mathbf M_3$ are the respective top-left and bottom-left $2 \times 2$ sub-matrices of the matricant $\mathbf{M}=\pmb{\mathcal{N}}\pmb{\mathcal{E}}\pmb{\mathcal{N}}^{-1}$. 
Here,
\begin{equation}\label{eq:matricant}
 \pmb{\mathcal{N}} = \begin{bmatrix}
\mathbf A & \overline{\mathbf A} \\ 
\mathbf B & \overline{\mathbf B}
\end{bmatrix}, \qquad \pmb{\mathcal{E}} = 
\text{Diag}\left(
\text e^{-2 kh}, 
\text e^{-2 (\lambda_1\lambda_2^{-1}) kh},
\text e^{2 kh},
\text e^{2 (\lambda_1\lambda_2^{-1}) kh} 
\right),
\end{equation} 
the overbar denotes the complex conjugate and $k$ is the wavenumber of the wrinkle. 
It is easy to check that $\det \mathbf z_{\mathrm{s}}=0$ is equivalent to the bifurcation criterion \eqref{eq:Flavin1}, as expected.
Also, notice that here the bifurcation criterion is independent of the material parameters $\beta_\mathrm{s}$, $\beta_\mathrm{f}$; only the stiffness ratio $\Gamma=\mu_\mathrm{f}/\mu_\mathrm{s}$ matters.

\begin{figure}[ht!]
\centering
\includegraphics[width=0.49\textwidth]{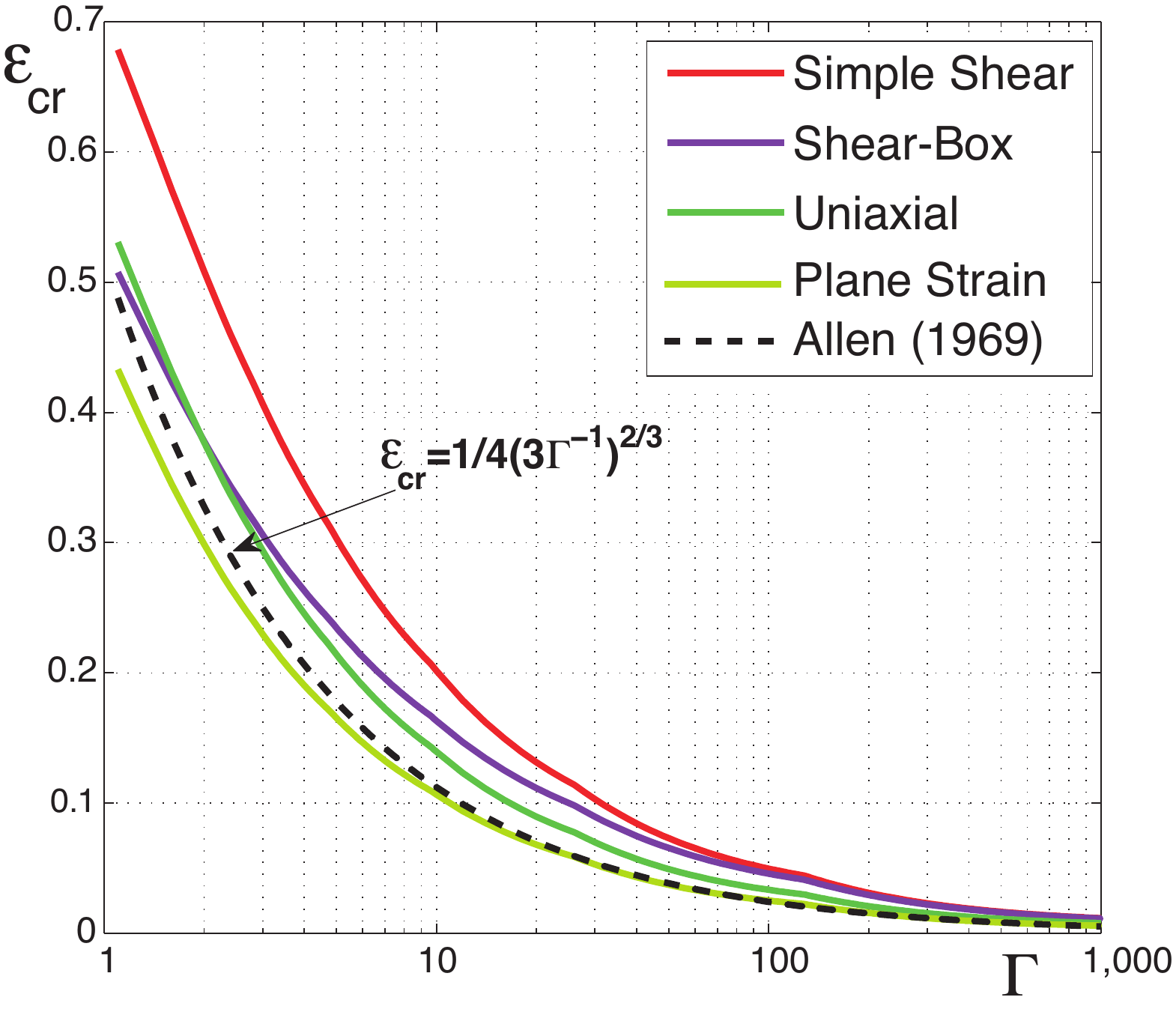} \hspace{-.3cm}
\raisebox{0.04\textwidth}{\includegraphics[width=0.52\textwidth]{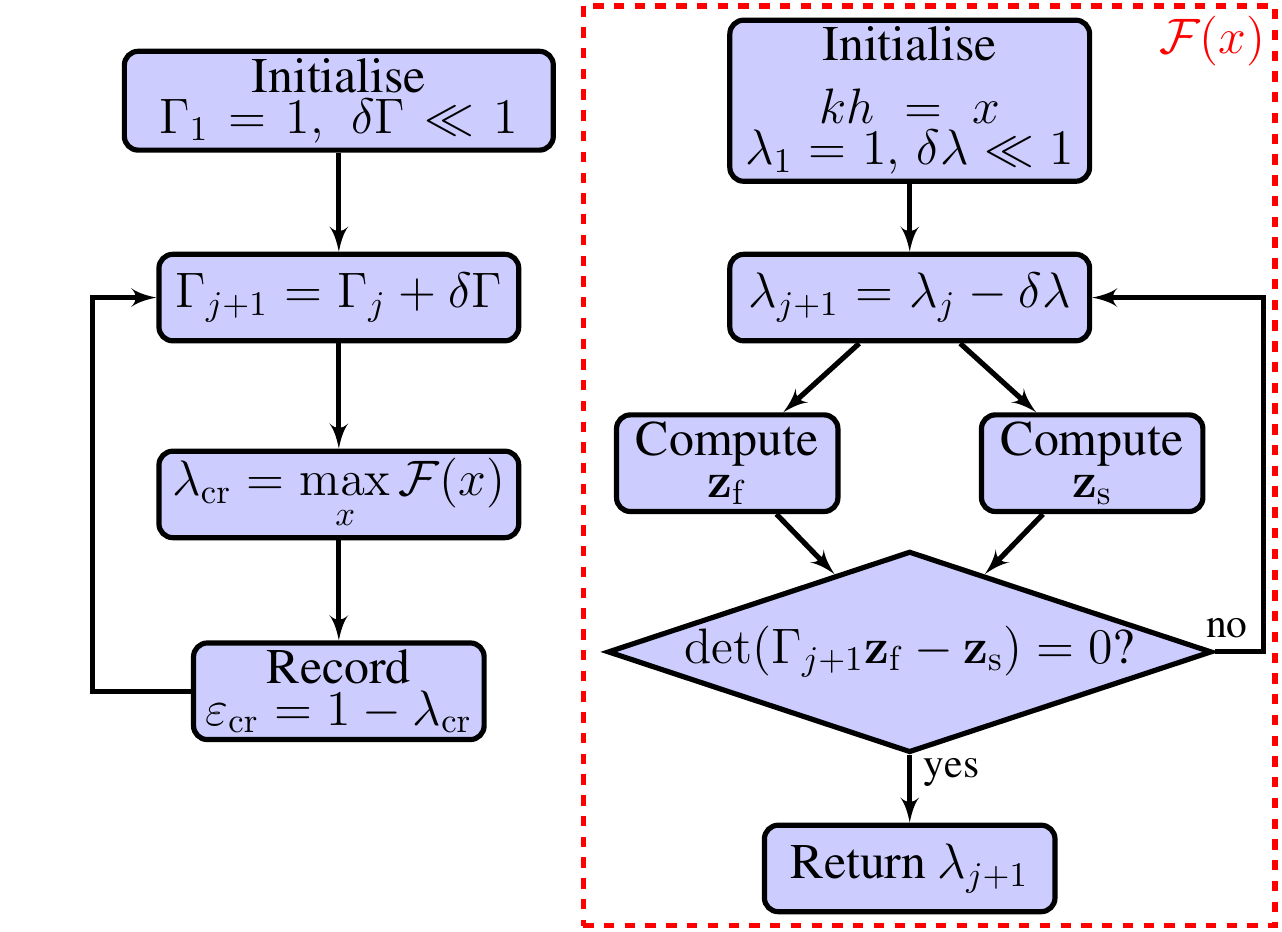}}
\caption{Left: Critical strain $\varepsilon_\mathrm{cr}$ for principal wrinkles on a coated substrate against the stiffness ratio $\Gamma$ for various compression tests. The result of Allen \cite{Allen} in linear elasticity is reported in dashed line. Right: Flowchart for the critical stretch finding routine (Here $x:=kh$ and $\mathcal{F}(x)$ denotes the algorithm in the red, dashed box).}
\label{fig:principal}
\end{figure}

We may then follow the same strategy as Cao and Hutchinson \cite{Huth_and_Cao_2012} to find the critical amount of deformation signalling the onset of sinusoidal wrinkles. 
When $\Gamma=1$, the system is homogeneous and non-dispersive: the wrinkles appear when $\lambda_1\lambda_2^{-1} = 0.2956$, see Equation (\ref{eq:Flavin1}). 
When $\Gamma>1$, the film/substrate structure is dispersive and to each value of $kh$ corresponds a value of $\lambda_1\lambda_2^{-1}$ for which Equation (\ref{crit}) is satisfied. 
By varying $kh$ we can find the maximal value for $\lambda_1\lambda_2^{-1}$, which will be the critical value at which the structure wrinkles.
Then by varying $\Gamma$, we can plot Figure \ref{fig:principal} (left), for all the considered cases of deformation. 
The routine described in the flowchart of Figure \ref{fig:principal} (right) allows us to obtain a general curve of the maximum critical stretch ratio $\lambda_1\lambda_2^{-1}$ against  $\Gamma$. 
Then, the critical value of $\lambda_1\lambda_2^{-1}$ is related to the critical value of stretch or strain according to each different mode of deformation Cases (i)-(iv).

Figure \ref{fig:principal} (left) recovers Figure 2 of Cao and Hutchinson \cite{Huth_and_Cao_2012} when the film and the substrate are both neo-Hookean and the deformation is plane strain.
In the figure we also report the analytical solution of Allen \cite{Allen} for the buckling of a thin strut on a half-space in linear elasticity. 
As pointed out in \cite{Huth_and_Cao_2012}, it becomes reliable when $\Gamma$ is large, i.e. when the deformation allowed before wrinkling is small and close to the linear regime.


\subsection{Oblique wrinkles}


For oblique wrinkles we must first integrate numerically the following differential Riccati equation for the film's surface impedance matrix $\mathbf z_{\mathrm{f}} = \mathbf z_{\mathrm{f}}(x_2)$,
\begin{equation}\label{eq:Riccati2}
\dfrac{\text d\mathbf z_{\mathrm{f}}}{\text dx_2} = \mathbf{z}_{\mathrm{f}} \mathbf{N}_2\mathbf{z}_{\mathrm{f}} - \text i\mathbf{z}_{\mathrm{f}} \mathbf{N}_1+ \text i\mathbf{N}^\dag_1\mathbf{z}_{\mathrm{f}}+\mathbf{N}_3,
\end{equation}
from its initial value \cite{ShEv02} $\mathbf z_{\mathrm{f}}(-h) = \mathbf 0$ to its value $\mathbf z_{\mathrm{f}}(h)$.
Then we compute the substrate's (constant) surface impedance matrix $\mathbf z_{\mathrm{s}}$ as the positive semi-definite solution to the algebraic Riccati equation \eqref{eq:Riccati1}. 
Finally we adjust the compressive stretch $\lambda$ until the following bifurcation criterion is met
\begin{equation} 
\text{det}\left(\mathbf z_{\mathrm{f}} - \mathbf z_{\mathrm{s}}\right) = 0.
\end{equation}
Notice  that here $\mathbf z_\mathrm{f}$ and $\mathbf z_\mathrm{s}$ are dimensional quantities, in contrast to the impedance matrices from Section~\ref{sec:Sec2}. because $\mathbf N_1$, $\mathbf N_2$, $\mathbf N_3$ in \eqref{eq:Riccati2} (in \eqref{eq:Riccati1}, respectively), depend on $\mu_\mathrm{f}$, $\beta_\mathrm{f} $ ($\mu_\mathrm{s}$, $\beta_\mathrm{s}$, respectively).

\begin{figure}[ht!]
\centering
\includegraphics[width=0.49\textwidth]{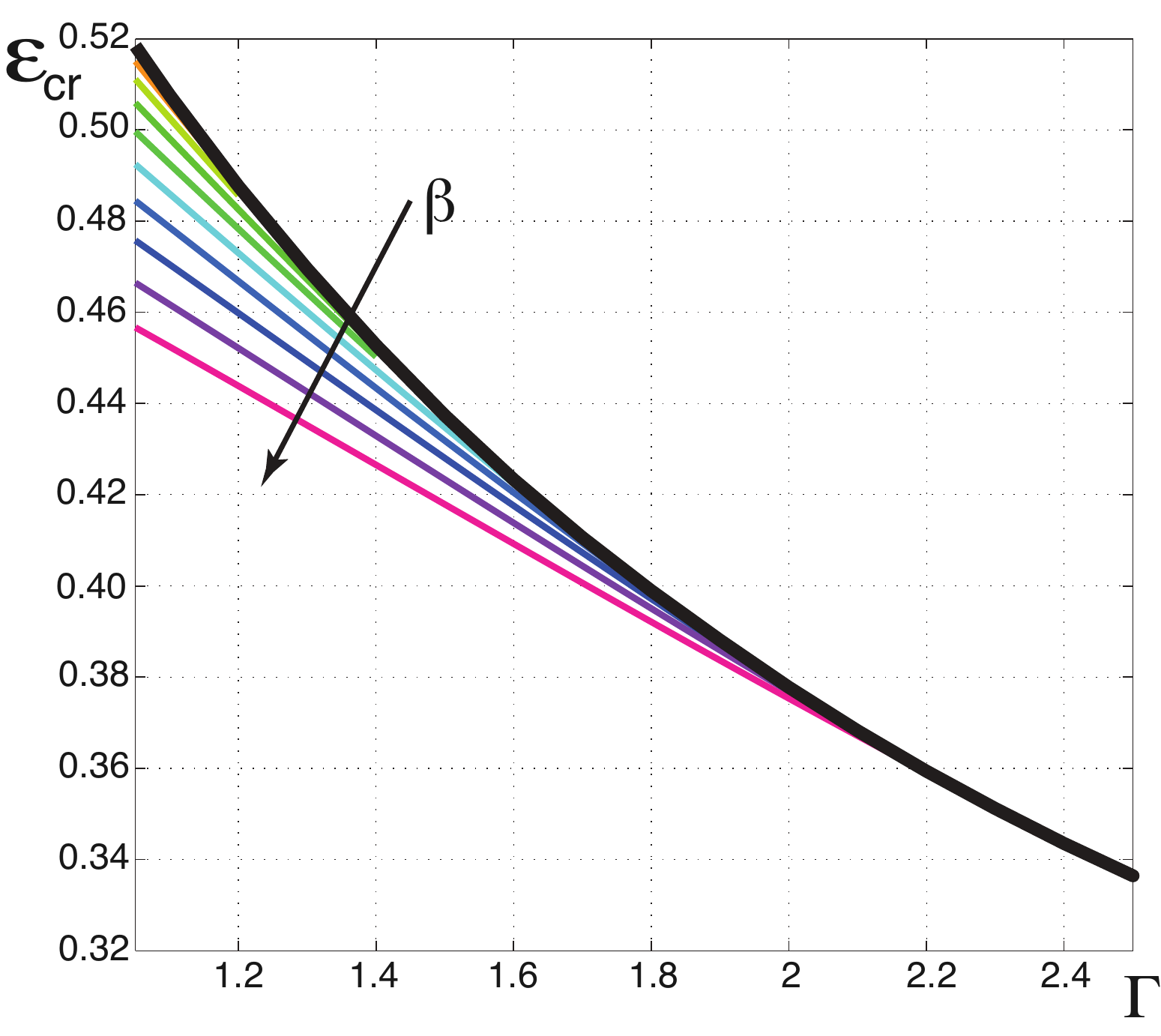} \hspace{-.6cm}
\raisebox{0.01\textwidth}{\includegraphics[width=0.53\textwidth]{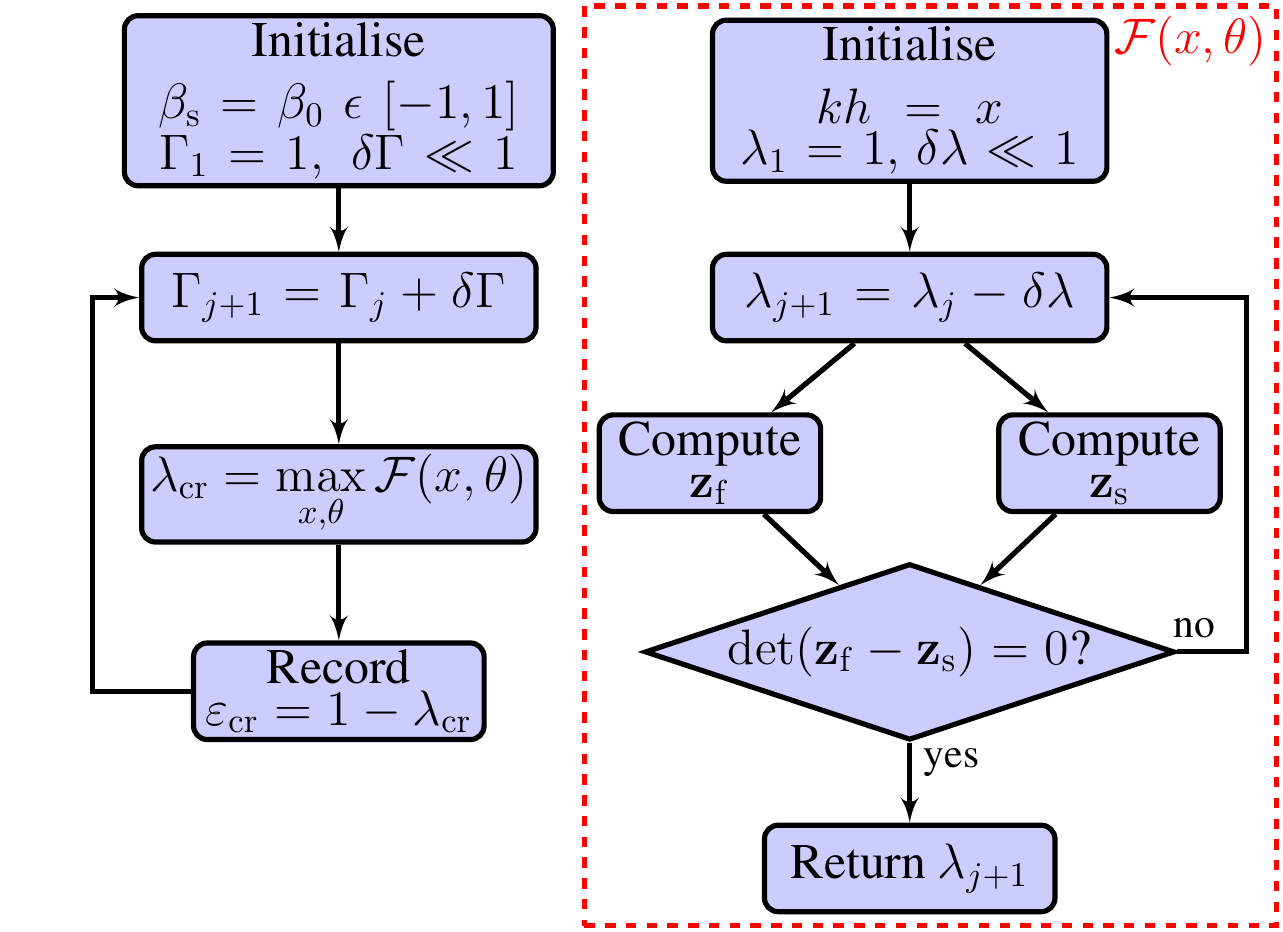}}
\caption{
Left: Critical strain $\varepsilon_\mathrm{cr}$ for oblique wrinkles on a coated substrate against the stiffness ratio $\Gamma$ for the shear-box compression test. 
The lower colour curves are for oblique wrinkles; they all merge into the upper black curve for principal wrinkles, early when $\beta_\mathrm{s}$ is away from $1.0$ and at about $\Gamma \simeq 2.3$ when it is close to $1$ (Extreme-Mooney model).
The coloured curves correspond to $\beta_\mathrm{s}=0.2,0.3,0.4,0.5,0.6,0.7,0.8,0.9,1.0$ as indicated by the arrow. For negative or null values of $\beta_\mathrm{s}$, only principal wrinkles are predicted as the favoured mode of buckling. 
Right: Flowchart for the critical stretch finding routine (Here $x:=kh$, and $\mathcal{F}(x,\theta)$ denotes the algorithm in the red, dashed box).}
\label{fig:oblique}
\end{figure}

Here the analysis is further complicated by the increase in the number of parameters.
For a given stiffness ratio $\Gamma$ and a given deformation, we must vary not only $kh$ to find the critical stretch, but also the obliquity angle $\theta$ and the material parameter $\beta_\mathrm{f}=\beta_\mathrm{s}$. 
Figure \ref{fig:oblique} (right) displays the flow-chart for the overall procedure used
(In particular we used the Matlab code \textit{fminsearch} to maximise the function $\mathcal{F}(kh,\theta)$.)
For our purposes it suffices to record, for a given deformation, the smallest critical strain $\varepsilon_\text{cr}$ obtained for an oblique wrinkle for a given $\beta_{\mathrm{f}}=\beta_{\mathrm{s}}$ and to plot that curve. Then by varying $\beta_{\mathrm{f}}=\beta_{\mathrm{f}}$ from $-1$ to $1$, we obtain a family of curves.
Here we present the results for the shear-box deformation, to compare them to our earlier experiments (we found similar figures for the other deformations).

Figure \ref{fig:oblique} (left) shows that oblique wrinkles (coloured curves underneath the thick black curve for principal wrinkles) precede principal wrinkles only when $\Gamma \le 2.3$ for the materials close to the Extreme-Mooney model ($\beta_\mathrm{s} \simeq 1.0$) and $\Gamma \le 2.0$ for most materials ($\beta_\mathrm{s}$ away from $1.0$).
(Specifically, we found that the Extreme-Mooney material predicts principal wrinkles when $\Gamma \ge 2.20, 1.50, 3.62, 2.28$ for Cases (i)$-$(iv), respectively.)

In conclusion: When the stiffness ratio $\Gamma$ is small, creases precede wrinkles; When it is large enough to support wrinkles, principal wrinkles precede oblique wrinkles. 
So, even though oblique wrinkles exist as first bifurcation mode in theory, they will never be observed experimentally. 

\color{black}


\section{Discussion}


To complete the paper we must qualify our concluding statements above.

First of all our stiffness ratio $\Gamma$ is one measure among others of the stiffness contrast between film and substrate. 
In practice it is unlikely that the proportion between $\mu_\text{0f}$ and $\mu_\text{0l}$, and $A_{\text f}$ and $A_{\text s}$, should be exactly the same, as in \eqref{stiffness-law}.
So a whole host of possibilities has been missed in this investigation.

Another caveat is that the Finite Element simulations of creases evoked in the paper all used the neo-Hookean strain energy density, which does not predict oblique wrinkles. 
Similarly, semi-analytical investigations into localised solutions, such as the one by Fu and Ciarletta \cite{FuCi15} also rely on neo-Hookean models.
It thus remains an open question whether oblique wrinkles can precede creases in simulations of compressed  Mooney-Rivlin solids. However, our study indicates that they could theoretically only appear in a narrow range of low stiffness contrast, where experiments (our table-top ones, and the comprehensive set of controlled ones presented by Jin et al. \cite{Jin15}) have so far failed to exhibit any evidence of their existence.

\color{black}
Also, guided by our experiments, we adopted the half-space approximation for the substrate, because the observed wavelength was small compared to the dimensions of the block. 
If the thickness $h$ of the stiff layer were comparable to that of the soft substrate, it might be possible that the finite depth of the substrate plays a role in the search for oblique wrinkles.
The competition between creases and wrinkles is also affected in that regime, and the onset of one compared to the other is more complicated than here, see the analysis of Jin et al. \cite{Jin15} for neo-Hookean solids.

\color{black}


\section*{Acknowledgments}

Part of this work is based on Chapter 1 of ALG's PhD thesis at NUI Galway, Ireland (2011-2015).
The financial support of the Hardiman Foundation (NUI Galway) and of the Government of Ireland Postgraduate Scholarship Scheme (Irish Research Council) for ALG and of Politecnico di Torino for MC is gratefully acknowledged.
We also thank Molly Aitken, Jorge Bruno, Juliet Destrade, Lucille Destrade, Sean Leen, Joanne McCarthy, Deirdre McGowan Smyth and Giuseppe Zurlo for their technical input.


\section*{Authors' contributions}
ALG carried out the experimental work and the calculations for surface instability, participated in the design of the study and drafted the manuscript.
MC extended the numerical calculations to the case of oblique wrinkles on a coated half-space, helped in drafting the manuscript, and provided flowcharts, schemes and figures coming out of numerical calculations.
MD conceived of the study, designed the study, coordinated the study and helped draft the manuscript. 
AG contributed with discussions about the study and helped in drafting the manuscript.
All authors gave final approval for publication.


\section*{Appendix}


The submatrices $\mathbf{N}_1$, $\mathbf{N}_2$, and $\mathbf{N}_3$ of the Stroh matrix can be found from the general expression in Ref.\cite{Destrade2005} as 
\begin{align*}
&\mathbf{N}_1=
\begin{bmatrix}
0&-\cos\theta&0\\
-\cos\theta&0 &-\sin\theta&\\
0&-\sin\theta&0\\
\end{bmatrix}, \quad  
\mathbf{N}_2=
\begin{bmatrix}
{1}/{\gamma_{21}}&0&0\\
0&0&0\\
0&0&{1}/{\gamma_{23}}
\end{bmatrix}, \\
& \mathbf{N}_3=
\begin{bmatrix}
-\eta&0&\kappa\\
0&-\nu&0\\
\kappa&0&-\psi
\end{bmatrix},
\end{align*}
where, for a Mooney Rivlin material \eqref{MR_Energy} or equivalently, a third-order elasticity solid \eqref{TO_Energy}, 
\begin{align*}
&\gamma_{ij}=\frac{\mu_0}{4}\left[\beta-1 + (\beta+1)\lambda_k^2\right]\lambda_i^2,
\\ 
& 
2\beta_{ij}=\gamma_{ij}+\gamma_{ji}, \qquad k\ne i,j,\\
& \psi =  \gamma_{13} \cos^2\theta + (\beta_{23} + \gamma_{23})\sin^2\theta, 
\\
&
\nu = (\gamma_{12}-\gamma_{21})\cos^2\theta + (\gamma_{32}-\gamma_{23})\sin^2\theta, \\
& \kappa = (\beta_{13}-\beta_{12}-\beta_{23}-\gamma_{21}-\gamma_{23})\cos\theta\sin\theta, 
\\ &
 \eta=  2 (\beta_{12}+\gamma_{21})\cos^2\theta + \gamma_{31} \sin^2\theta.
\end{align*}


\end{document}